\documentclass[aps,prl,twocolumn,groupedaddress,showpacs]{revtex4}
\usepackage{bm}
\usepackage{graphicx}
\begin{document}

\title{Angle-Resolved Photoemission Spectroscopy of Iron-Chalcogenide Superconductor Fe$_{1.03}$Te$_{0.7}$Se$_{0.3}$ : Strong-Coupling Superconductivity and Universality of Inter-Band Scattering}

\author{
	K. Nakayama,$^1$
	T. Sato,$^{1,2}$
	P. Richard,$^3$
	T. Kawahara,$^1$
	Y. Sekiba,$^1$
	T. Qian,$^1$
	G. F. Chen,$^4$
	J. L. Luo,$^4$
	N. L. Wang,$^4$
	H. Ding,$^4$
	and T. Takahashi$^{1,3}$}

\affiliation{$^1$Department of Physics, Tohoku University, Sendai 980-8578, Japan}
\affiliation{$^2$TRiP, Japan Science and Technology Agency (JST), Kawaguchi 332-0012, Japan}
\affiliation{$^3$WPI Research Center, Advanced Institute for Materials Research, Tohoku University, Sendai 980-8577, Japan}
\affiliation{$^4$Beijing National Laboratory for Condensed Matter Physics, and Institute of Physics, Chinese Academy of Sciences, Beijing 100190, China}

\date{\today}

%\begin{minipage}[t]{6.8in}
\begin{abstract}
We have performed high-resolution angle-resolved photoemission spectroscopy of iron-chalcogenide superconductor Fe$_{1.03}$Te$_{0.7}$Se$_{0.3}$ ($T_c$ = 13 K) to investigate the electronic structure relevant to superconductivity.  We observed a hole- and an electron-like Fermi surfaces at the Brillouin zone center and corner, respectively, which are nearly nested by the $Q$$\sim$($\pi$, $\pi$) wave vector. We do not find evidence for the nesting instability with $Q$$\sim$($\pi$+$\delta$, 0) reminiscent of the antiferromagnetic order in the parent compound Fe$_{1+y}$Te. We have observed an isotropic superconducting gap along the hole-like Fermi surface with the gap size $\Delta$ of $\sim$4 meV (2$\Delta$/$k_BT_c$$\sim$7), demonstrating the strong-coupling nature of the superconductivity.  The observed similarity of low-energy electronic excitations between iron-chalcogenide and iron-arsenide superconductors strongly suggests that common interactions which involve $Q$$\sim$($\pi$, $\pi$) scattering are responsible for the superconducting pairing.
\end{abstract}

\pacs{74.25.Jb, 74.70.-b, 79.60.-i}

%\end{minipage}
\maketitle
%\narrowtext

The discovery of new iron-based superconductors \cite{Kamihara, Chen, AIST, Ren, Wang} with the highest critical temperature ($T_c$) of $\sim$55 K has triggered fierce debates on the superconducting mechanism, since it is hard to explain such a high-$T_c$ value within the conventional phonon-mediated superconducting framework \cite{Boeri}.  Experimentally, both hole- and electron-doped BaFe$_2$As$_2$ superconductors ({\it 122} system) exhibit an anomalously strong pairing behavior on small Fermi surfaces (FSs) which are connected by the $Q$$\sim$($\pi$, $\pi$) wave vector, suggesting the importance of inter-band interactions for the occurrence of high-$T_c$ superconductivity \cite{HongGap, Terashima}.  Since the parent compounds of the {\it 122} system commonly show an antiferromagnetic (AF) long-range order with the $Q$=($\pi$, $\pi$) propagation vector \cite{Huang}, remnant AF spin fluctuations with a similar wave vector may be responsible for the pairing interactions in doped compounds \cite{Mazin, Kuroki, Lee, Hu, Yao, Tesanovic}.  However, the universality of the inter-band scattering $via$ $Q$$\sim$($\pi$, $\pi$) in all the iron-based superconductors as well as the role of magnetism for the pairing are still unclear.  In particular, the FS topology of the iron-chalcogenide superconductor Fe$_{1+y}$Te$_{1-x}$Se$_x$ ({\it 11} system) \cite{Hsu, Yeh, Sales, Fang, Mizuguchi, GFChen} is highly controversial.  Although the density functional calculations for stoichiometric FeTe(Se) have predicted that the FS shape is very similar to that of FeAs compounds \cite{Subedi}, a remarkably different FS may emerge in the actual case due to an electron doping through the introduction of excess Fe atoms \cite{Zhang, Han}.  In addition, the appearance of AF order with $Q$=($\pi$+$\delta$, 0) in the parent compound Fe$_{1+y}$Te \cite{Bao, DaiFeTe}, corresponding to a 45$^{\circ}$-rotated AF order as compared to that of the {\it 122} system, raises a major question concerning the relationship between magnetism and superconductivity.  The presence of multiple spin excitations in the Se-substituted superconducting samples makes this issue even more complicated \cite{Bao, DaiFeTe, Mook, Iikubo, Qiu}.  It is thus definitely important to clarify the low-energy electronic excitations in iron-chalcogenide superconductor Fe$_{1+y}$Te$_{1-x}$Se$_x$ and compare their relationship with the magnetic excitations observed by neutron scattering.

In this Letter, we report results of high-resolution angle-resolved photoemission spectroscopy (ARPES) of Fe$_{1.03}$Te$_{0.7}$Se$_{0.3}$ superconductor ($T_c$ = 13 K).  We show that the FS topology of this superconductor is essentially similar to that of the optimally-doped BaFe$_2$As$_2$ \cite{HongGap, Terashima}, in contrast to a recent ARPES study of the parent compound Fe$_{1+y}$Te, where an additional FS is found at the X point \cite{Hasan}.  The observed FS topology suggests the importance of inter-band interactions $via$ $Q$$\sim$($\pi$, $\pi$) wave vector for the pairing.  In addition, we observed an isotropic superconducting gap with a large 2$\Delta$/$k_BT_c$ ratio on the hole FS.  We discuss the interplay between the AF fluctuations and the occurrence of superconductivity.

The high-quality single crystals of Fe$_{1.03}$Te$_{0.7}$Se$_{0.3}$ ($T_c$ = 13 K) used in this study were grown by the Bridgeman method \cite{GFChen}.  High-resolution ARPES measurements were performed at beamline 28A in Photon Factory, KEK, Tsukuba, using circularly polarized photons ($h\nu$ = 44 and 100 eV) and at Tohoku University using the He I$\alpha$ resonance line ($h\nu$ = 21.218 eV).  The energy resolution was set at 1.7, 12, and 35 meV for the measurements of the superconducting gap, the valence band dispersion, and the core levels, respectively.  The angular resolution was set at 0.2$^{\circ}$.  Since the sample is unstable in the atmosphere \cite{Xia}, almost all of the preparation procedures such as sample mounting were done in a glove box filled with Ar gas.  Clean surfaces for the ARPES measurements were obtained by cleaving crystals $in$ $situ$ in a working vacuum better than 1$\times$10$^{-10}$ Torr.  The Fermi level ($E_F$) of the samples was referenced to that of a gold film evaporated onto the sample substrate.  Low-energy electron diffraction from the measured surface shows a sharp 1$\times$1 pattern without any detectable reconstruction down to 80 K.

\begin{figure}[!t]
\begin{center}
\includegraphics[width=2.5in]{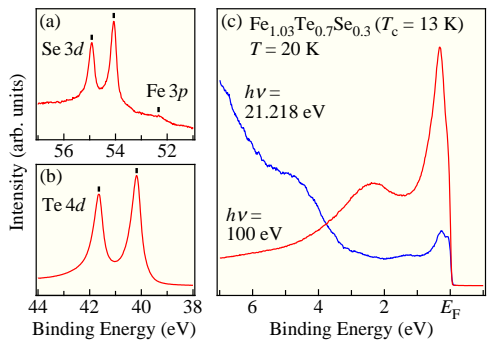}
\end{center}
\caption{
(color online) Core level photoemission spectrum of Fe$_{1.03}$Te$_{0.7}$Se$_{0.3}$ ($T_c$ = 13 K) for (a) Se 3$d$ and Fe 3$p$ and (b) Te 4$d$, measured with 100 eV photons at $T$ = 20 K.  The energy position of each core level is indicated by a black bar.  (c) Photoemission spectra in the valence band region measured with $h\nu$ = 100 eV (red curve) and the He I$\alpha$ line ($h\nu$ = 21.218 eV: blue curve).
}
\end{figure}

Figure 1(a) shows the Se 3$d$ and Fe 3$p$ core level spectrum of Fe$_{1.03}$Te$_{0.7}$Se$_{0.3}$ ($T_c$ = 13 K).  Two sharp peaks at the binding energies of 54.1 and 54.9 eV are ascribed to the spin-orbit split Se 3$d_{5/2}$ and Se 3$d_{3/2}$ core levels.  It has been reported that Se 3$d$ core levels appear at ~53.8 and ~54.6 eV on hexagonal non-superconducting FeSe$_{1-x}$ \cite{Yamasaki}.  We do not find any of these peaks in the present result, suggesting the absence of such an impurity phase.  As shown in Fig. 1(b), the Te 4$d$ core levels (40.2 and 41.6 eV) exhibit well defined Lorentzian-like peaks with small asymmetry, indicating the absence of amorphous Te phase \cite{Xia} and no significant surface contamination.  To clarify the character of the valence band, we preformed measurements with two different photon energies.  As shown in Fig. 1(c), we observed pronounced peaks within $\sim$3 eV below $E_F$, whose intensity is significantly enhanced at $h\nu$ = 100 eV.  On the other hand, the peak around $\sim$4.5 eV looks pronounced at $h\nu$ = 21.218 eV.  Since the photo-ionization cross-section ratio of the Fe 3$d$ orbitals to the Te 5$p$ (Se 4$p$) orbitals is $\sim$50 ($\sim$40) and $\sim$1 ($\sim$0.6) for 100 eV and 21.218 eV photons, respectively \cite{cross-section}, we attribute the near-$E_F$ peaks and the 4.5 eV peak to Fe-3$d$ and Te-5$p$ (Se-4$p$) dominant states, respectively.   This assignment is consistent with previous angle-integrated photoemission results on FeSe$_{1-x}$ \cite{Yamasaki, Yoshida}.

\begin{figure}[!t]
\begin{center}
\includegraphics[width=3.4in]{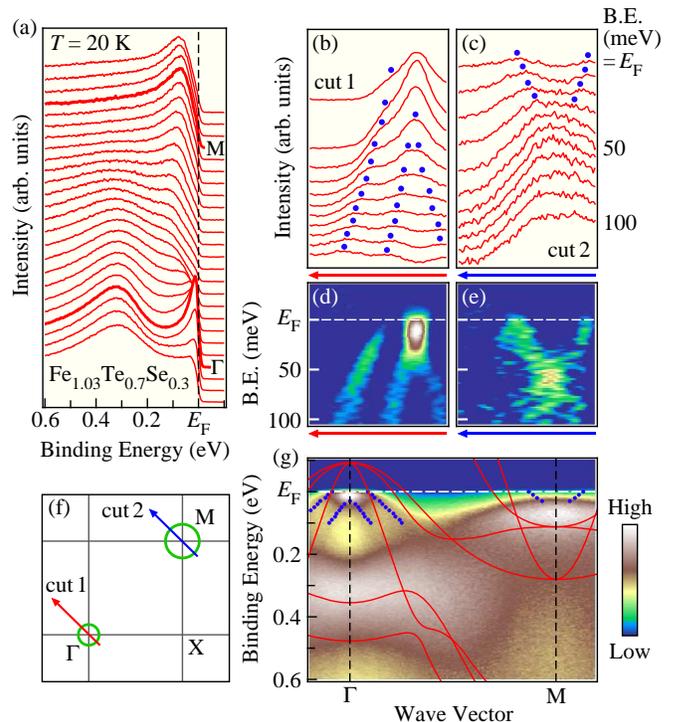}
\end{center}
\caption{
(color online) (a) ARPES spectra near $E_F$ of Fe$_{1.03}$Te$_{0.7}$Se$_{0.3}$ measured along the $\Gamma$-M high-symmetry line at 20 K with $h\nu$ = 44 eV.  (b) and (c) Representative MDCs in the vicinity of $E_F$ measured along cut 1 and 2 in (f), respectively.  Blue dots denote the peak positions of MDCs.  (d) and (e) Second derivative plot of MDCs along cut 1 and 2, respectively, as a function of binding energy and wave vector.  (f) Schematic FS (green curves) with the locations of momentum cuts (red and blue arrows).  (g) Intensity plot of (a) as a function of binding energy and wave vector together with the calculated energy bands for FeTe at $k_z$ = 0 (red curves) \cite{Subedi}.  Experimental near-$E_F$ band dispersions extracted from the MDC peak positions are also shown by blue dots.
}
\end{figure}

Figure 2(a) shows the ARPES spectra measured along the $\Gamma$-M high-symmetry line of the Brillouin zone for the two-Fe unit cell.  A hole-like band is clearly observed, and appears to cross $E_F$ around the $\Gamma$ point.  We also observe a less dispersive band at $\sim$100 meV around the M point and a broad feature at $\sim$300 meV around the $\Gamma$ point.  To see the near-$E_F$ dispersion in more detail, we show in Figs. 2(b) and (c) the momentum distribution curves (MDCs) and their second derivative intensities [Figs. 2(d) and (e)], measured along cuts crossing the $\Gamma$ and M points, respectively [cuts 1 and 2 in Fig. 2(f)].  As clearly seen in Figs. 2(b) and (d), there are two hole-like bands centered at the $\Gamma$ point.  The outer band creates a hole FS pocket, whereas the inner band does not cross $E_F$.  Around the M point [Figs. 2(c) and (e)], we observe a shallow electron-like band crossing $E_F$, whose bottom of dispersion is at $\sim$50 meV.  The estimated Fermi velocity is $\sim$0.4 eV\AA \  for both the hole and the electron bands, comparable to that for optimally-doped {\it 122} compound \cite{HongBand}.  In Fig. 2(g), we plot the ARPES intensity compared with the density functional calculations for stoichiometric FeTe (red curves) \cite{Subedi}.  The calculated bands are renormalized by a factor of 2.  The experimental low-energy band dispersions extracted from the peak positions of MDCs are also plotted as blue dots for comparison.  Although the observed overall band structure appears to roughly track the renormalized band calculations, there are remarkable discrepancies especially in the near-$E_F$ region.  For instance, the observed two hole-like bands at the $\Gamma$ point are located at higher binding energy as compared to the calculated bands.  On the other hand, the observed electron pocket at the M point is located at lower binding energy.  This opposite shift between the hole and the electron pockets cannot be simply explained in terms of electron doping by the presence of excess Fe atoms.   Interestingly, a similar behavior has been reported for FeP (iron-phosphorous)- \cite{dHvA} and FeAs-based \cite{HongBand, Shen} superconductors, suggesting that it is a general trend of Fe-based superconductors.

\begin{figure}[!t]
\begin{center}
\includegraphics[width=2.5in]{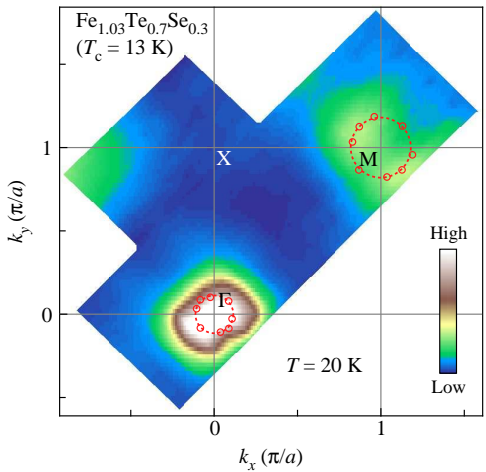}
\end{center}
\caption{
(color online) ARPES intensity plot at $E_F$ of Fe$_{1.03}$Te$_{0.7}$Se$_{0.3}$ as a function of the two-dimensional wave vector measured at 20 K with 44 eV photons.  The intensity at $E_F$ is obtained by integrating the spectra within $\pm$10 meV with respect to $E_F$.  Solid and dashed red circles show experimentally determined $k_F$ points and schematic FSs, respectively.  There are sizable experimental uncertainties on the experimentally determined $k_F$ points, mainly due to weak intensity around the M point.
}
\end{figure}

In Fig. 3, we plot the ARPES intensity at $E_F$ as a function of the two-dimensional wave vector.  We clearly identify FSs centered at the $\Gamma$ and the M points, corresponding to the hole and the electron pockets, respectively [see also Figs. 2(d) and (e)], essentially similar to the FS topology of the {\it 122} superconductors \cite{HongGap, Terashima, Nakayama}.  No indication of FS is found at the X point, in contrast to a recent ARPES study of the parent compound Fe$_{1+y}$Te \cite{Hasan}.  The disappearance of the FS around X is consistent with the disappearance of long-range AF order with the AF wavevector of ($\pi$, 0) in the superconducting sample, which would fold the FS at $\Gamma$ to the X point in the parent compound.

\begin{figure}[!t]
\begin{center}
\includegraphics[width=3.4in]{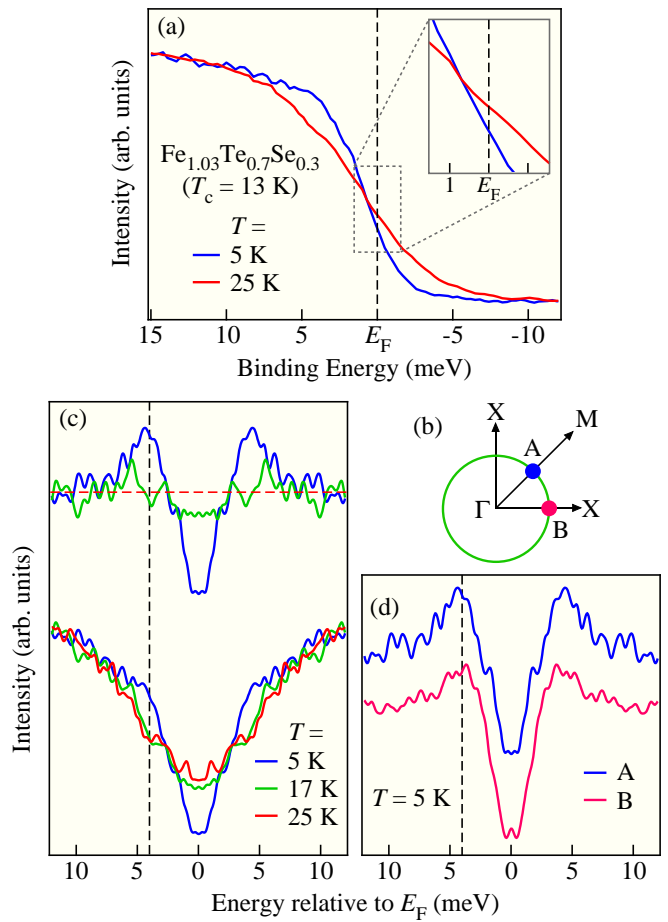}
\end{center}
\caption{
(color online) (a) Ultrahigh-resolution ARPES spectra of Fe$_{1.03}$Te$_{0.7}$Se$_{0.3}$ near $E_F$ at $T$ = 5 and 25 K (blue and red curve, respectively) with the He I$\alpha$ resonance line, measured at point A of the outer hole pocket displayed in (b).  The inset shows the expansion in the vicinity of $E_F$.  (b) Schematic hole-like FS at the $\Gamma$ point with the location of the $k_F$ points A and B.  (c) Temperature dependence of symmetrized ARPES spectra (bottom) at point A, and the same but divided by the spectrum at $T$ = 25 K (top).  (d) Comparison of the symmetrized spectra at $k_F$ points along $\Gamma$-M (blue curve) and $\Gamma$-X (purple curve) high-symmetry lines measured at 5 K divided by the 25 K spectrum.  Dashed lines at 4 meV in (c) and (d) represent the energy scale of the superconducting gap $\Delta$.
}
\end{figure}

In order to elucidate the character of the superconducting gap, we have performed ultrahigh-resolution ($\Delta$$E$ = 1.7 meV) ARPES measurements in the close vicinity of $E_F$.  Figure 4(a) shows the temperature dependence of the near-$E_F$ ARPES spectrum measured across $T_c$ at $k_F$ of the outer hole-like band as denoted by a blue dot in Fig. 4(b) \cite{RefMpoint}.  As seen in Fig. 4(a), the midpoint of the leading edge at 5 K is shifted toward higher binding energy by $\sim$0.6 meV, suggesting the opening of a superconducting gap, and the intersection of the ARPES spectra measured above and below $T_c$ is also away from $E_F$ (see inset).  To eliminate the effect of the Fermi-Dirac distribution function, we have symmetrized the ARPES spectrum at each temperature as shown in the bottom of Fig. 4(c).  On decreasing temperature, the spectral weight near $E_F$ is transferred to high binding energy below $T_c$.  The spectral shape is distinctly different between 5 K and 17 K, whereas no essential difference is seen between 17 K and 25 K, indicating that the spectral feature varies drastically across $T_c$.  We divided the symmetrized ARPES spectra measured at 5 and 17 K by the 25 K spectrum [top of Fig. 4(c)] to cancel out the V shaped spectral density of states due to the tail of the inner hole-like band.  Apparently, a coherence peak emerges at the binding energy of $\sim$4 meV at 5 K, unambiguously demonstrating the opening of a superconducting gap.  The estimated superconducting gap size ($\Delta$$\sim$4 meV) corresponds to a 2$\Delta$/$k_BT_c$ value of $\sim$7, indicating the strong-coupling nature of superconductivity in Fe$_{1.03}$Te$_{0.7}$Se$_{0.3}$.  To examine a possible gap anisotropy, we measured an ARPES spectrum along the $\Gamma$-X cut [point B in Fig. 4(b)], and show the symmetrized $k_F$ spectrum at 5 K after division by the 25 K spectrum together with that along the $\Gamma$-M cut (point A).  We immediately notice that the observed energy position of the coherence peak for points A and B almost coincides, suggesting that the superconducting gap is likely to be isotropic within the present experimental accuracy.

Now we discuss the implication of the present experimental results in comparison with the {\it 122} compounds.  It is useful to summarize the present ARPES results: (i) we observed a hole pocket at the $\Gamma$ point and an electron pocket at the M point, (ii) the hole pocket at the X point observed in Fe$_{1+y}$Te \cite{Hasan} is absent in Fe$_{1.03}$Te$_{0.7}$Se$_{0.3}$, and (iii) a large and nearly-isotropic superconducting gap ($\Delta$$\sim$4 meV) with a strong-coupling value (2$\Delta$/$k_BT_c$$\sim$7) opens at the hole pocket.  These experimental facts show that the low-energy electronic structure responsible for the superconductivity in Fe$_{1.03}$Te$_{0.7}$Se$_{0.3}$ is qualitatively similar to that of the {\it 122} superconductors \cite{HongGap, Terashima, Nakayama}, indicating that the effect of electron doping by an excess of Fe is overestimated in a recent theoretical study which predicted the appearance of a large square-type FS both at the $\Gamma$ and the X points due to the chemical potential shift \cite{Han}.  The present results further suggest that the $Q$$\sim$($\pi$+$\delta$, 0) spin fluctuations observed by the inelastic neutron scattering of the Se-substituted superconducting sample \cite{Bao, DaiFeTe, Iikubo} are less important for the superconducting pairing in Fe$_{1.03}$Te$_{0.7}$Se$_{0.3}$, since the $Q$$\sim$($\pi$+$\delta$, 0) wave vector does not connect the observed FSs as clearly seen in Fig. 3.  The observed FS topology and the strong-coupling superconducting gap parameter in Fe$_{1.03}$Te$_{0.7}$Se$_{0.3}$ favors the scenario that a coupling of electrons to excitations near $Q$=($\pi$, $\pi$) promotes the inter-band scattering between the hole and the electron pockets, leading to a strong superconducting pairing \cite{HongGap, Terashima, Nakayama}.  This suggests the universality of pairing interactions between {\it 122} and {\it 11} systems.  Interestingly, recent neutron scattering measurements of superconducting Fe$_{1+y}$Te$_{1-x}$Se$_x$ reported the presence of spin fluctuations near ($\pi$, $\pi$) even well above $T_c$ \cite{Mook, Iikubo, Qiu}, consistent with the inter-band scattering condition observed in the present ARPES study.  It is thus inferred that the inter-band interactions are magnetic in origin, and as a consequence, an unconventional spin-mediated pairing emerges in both {\it 122} and {\it 11} systems.

Finally, we briefly mention the relationship between the present observations and the recent experimental results on the composition dependence of physical properties in Fe$_{1+y}$Te$_{1-x}$Se$_x$.  It has been reported that the superconducting transition of Fe$_{1+y}$Te$_{1-x}$Se$_x$ is better defined ($e.g.$ a clear specific heat jump at $T_c$) in crystals with a smaller amount of excess Fe ($y$) and a larger Se concentration ($x$) up to $x$ = 0.5 \cite{Sales, Liu}.  We think that the likely competing  $Q$$\sim$($\pi$+$\delta$, 0) and the $Q$$\sim$($\pi$, $\pi$) spin fluctuations play different roles to the superconductivity and that this competition would be a key to understand the composition dependence of the superconducting character.  For samples with a larger excess of Fe ($y$) and smaller Se values ($x$), the presence of ($\pi$+$\delta$, 0) spin excitations \cite{Bao, DaiFeTe, Iikubo} may cause a reduction of the superconducting volume fraction, since those spin fluctuations would not be favorable to the superconducting pairing as discussed above.  With decreasing excess Fe and increasing Se concentration, the ($\pi$+$\delta$, 0) spin fluctuations are suppressed, whereas the ($\pi$, $\pi$) spin fluctuations, which would assist the pairing and lead to stronger superconductivity, are enhanced, in agreement with the inelastic neutron scattering measurements \cite{Bao, DaiFeTe, Mook, Iikubo, Qiu}.  A systematic study of the electronic structure for a wide range of excess Fe ($x$) and Se compositions ($y$) is the next step toward a full understanding of the superconducting mechanism in Fe$_{1+y}$Te$_{1-x}$Se$_x$.

In conclusion, we have studied the shallow core levels, the band dispersions, the FS topology and the superconducting gap of Fe$_{1.03}$Te$_{0.7}$Se$_{0.3}$ by using high-resolution ARPES.  While the experimentally determined low-energy electronic structure is distinctly different from that of the parent AF phase, we found that it is qualitatively similar to that of the {\it 122} superconductors.  Our results suggest that the coupling to $Q$=($\pi$+$\delta$, 0) AF correlations is suppressed in Fe$_{1.03}$Te$_{0.7}$Se$_{0.3}$ and the superconducting state arises from the inter-band interactions between hole and electron FSs $via$ $Q$$\sim$($\pi$, $\pi$), suggesting the importance of common spin fluctuations to the superconducting pairing of Fe-based superconductors.

We thank M. Kubota and K. Ono for their help in the experiment. This work was supported by grants from JSPS, TRiP-JST, MEXT of Japan, CREST-JST, the Chinese Academy of Sciences, NSF, Ministry of Science and Technology of China, and NSF of US.  ARPES measurements in synchrotron were carried out at KEK-PF (Proposal No. 2009S2-005).  K.N. and T.Q. thank JSPS for financial support.


\begin{thebibliography}{99}

\bibitem{Kamihara} Y. Kamihara $et$ $al$., J. Am. Chem. Soc. {\bf 130}, 3296 (2008).
\bibitem{Chen} X. H. Chen $et$ $al$., Nature (London) {\bf 453}, 1224 (2008).
\bibitem{AIST} H. Kito $et$ $al$., J. Phys. Soc. Jpn. {\bf 77}, 063707 (2008).
\bibitem{Ren} Z. A. Ren $et$ $al$., Chin. Phys. Lett. {\bf 25}, 2215 (2008).
\bibitem{Wang} C. Wang $et$ $al$., Europhys. Lett. {\bf 83}, 67006 (2008).
\bibitem{Boeri} L. Boeri $et$ $al$., Phys. Rev. Lett. {\bf 101}, 026403 (2008).
\bibitem{HongGap} H. Ding $et$ $al$., Europhys. Lett. {\bf 83}, 47001 (2008).
\bibitem{Terashima} K. Terashima $et$ $al$., Proc. Natl. Acad. Sci. USA {\bf 106}, 7330 (2009).
\bibitem{Huang} Q. Huang $et$ $al$., Phys. Rev. Lett. {\bf 101}, 257003 (2008).
\bibitem{Mazin} I. I. Mazin $et$ $al$., Phys. Rev. Lett. {\bf 101}, 057003 (2008).
\bibitem{Kuroki} K. Kuroki $et$ $al$., Phys. Rev. Lett. {\bf 101}, 087004 (2008).
\bibitem{Lee} F. Wang $et$ $al$., Phys. Rev. Lett. {\bf 102}, 047005 (2009).
\bibitem{Hu} K. Seo $et$ $al$., Phys. Rev. Lett. {\bf 101}, 206404 (2008).
\bibitem{Yao} Z. J. Yao, J. X. Li and Z. D. Wang, New J. Phys. {\bf 11}, 025009 (2009).
\bibitem{Tesanovic} V. Cvetkovic and Z. Tesanovic, Europhys. Lett. {\bf 85}, 37002 (2009).
\bibitem{Hsu} F. C. Hsu $et$ $al$., Proc. Natl. Acad. Sci. USA {\bf 105}, 14262 (2008).
\bibitem{Yeh} K. W. Yeh $et$ $al$., Europhys. Lett. {\bf 84}, 37002 (2008).
\bibitem{Sales} B. C. Sales $et$ $al$., Phys. Rev. B {\bf 79}, 094521 (2009).
\bibitem{Fang} M. H. Fang $et$ $al$., Phys. Rev. B {\bf 78}, 224503 (2008).
\bibitem{Mizuguchi} Y. Mizuguchi $et$ $al$., arXiv:0811.1123 (2008).
\bibitem{GFChen} G. F. Chen $et$ $al$., Phys. Rev. B {\bf 79}, 140509(R) (2009).
\bibitem{Subedi} A. Subedi $et$ $al$., Phys. Rev. B {\bf 78}, 134514 (2008).
\bibitem{Zhang} L. Zhang $et$ $al$., Phys. Rev. B {\bf 79}, 012506 (2009).
\bibitem{Han} M. J. Han and S. Y. Savrasov, arXiv:0903.2896 (2009).
\bibitem{Bao} W. Bao $et$ $al$., Phys. Rev. Lett. {\bf 102}, 247001 (2009).
\bibitem{DaiFeTe} S. Li $et$ $al$., Phys. Rev. B {\bf 79}, 054503 (2009).
\bibitem{Mook} H. A. Mook $et$ $al$., arXiv:0904.2178 (2009).
\bibitem{Iikubo} S. Iikubo $et$ $al$., arXiv:0904.3824 (2009).
\bibitem{Qiu} Y. Qiu $et$ $al$., arXiv:0905.3559 (2009).
\bibitem{Hasan} Y. Xia $et$ $al$., arXiv:0901.1299 (2009).
\bibitem{Xia} T.-L. Xia $et$ $al$., Phys. Rev. B {\bf 79}, 140510(R) (2009).
\bibitem{Yamasaki} A. Yamasaki $et$ $al$., arXiv:0902.3314 (2009).
\bibitem{cross-section} J. J. Yeh and I. Lindau, Atomic Data and Nuclear Tables {\bf 32}, 1 (1985).
\bibitem{Yoshida} R. Yoshida $et$ $al$., J. Phys. Soc. Jpn. {\bf 78}, 034708 (2009).
\bibitem{HongBand} H. Ding $et$ $al$., arXiv:0812.0534 (2008).
\bibitem{dHvA} A. I. Coldea $et$ $al$., Phys. Rev. Lett. {\bf 101}, 216402 (2008).
\bibitem{Shen} M. Yi $et$ $al$., arXiv:0902.2628 (2009).
\bibitem{Nakayama} K. Nakayama $et$ $al$., Europhys. Lett. {\bf 85}, 67002 (2009).
\bibitem{RefMpoint} It is difficult to obtain reliable superconducting gap data on the electron pocket due to fairly weak photoelectron intensity near $E_F$.
\bibitem{Liu} T. J. Liu $et$ $al$., arXiv:0904.0824 (2009).

\end{thebibliography}
\end{document}